\documentstyle[12pt]{article}
\normalbaselines
\begin{document}

\begin{center}
{\large\bf "MINUS C" SYMMETRY IN CLASSICAL AND QUANTUM THEORIES}\\ [1 mm]

G.A. Kotel'nikov\\ { \it Russian Research Center " Kurchatov Institute ",
Moscow, 123182 Russia}\\ e-mail: kga@kga.kiae.su\\ [ 2 mm ]
\end{center}

\begin{abstract}
It is shown that the transformations  of  the  charge  conjugation  in
classical  electrodynamics and in quantum theory can be interpreted as
the consequences of the symmetry of Maxwell and Dirac  equations  with
respect  to the inversion of the speed of light $t \to t,  {\bf x} \to
{\bf x},  \ c \to -c$.  The elements of  physical  interpretation  are
given. \end{abstract}

\section{Introduction}
\label{s1}
\par A symmetry approach plays a large part  in  theoretical  physics.
Below   we   call  attention  to  the  discrete  symmetries.  Discrete
transformations are well known in particle  physics  \cite{Sch57},  in
quantum field theory \cite{Ber68}, in nuclear physics \cite{Shi72}, in
quantum  physics  by  studying  systems   with   periodic     changing
parameters \cite{Mal79}. The space inversion $P({\bf x} \to -{\bf x})$
reflecting the right - left symmetry of  the  3-dimensional  space  of
events,  the  time  reversal  $T(t  \to  -t)$  describing the symmetry
connected with change of sign of  time,  the  charge  conjugation  $C$
connected with symmetry of particle - antiparticle,  the wave function
transformations    of    the    type    $\Psi_{\varepsilon}     (t+T)=
e^{-i\varepsilon  T/\hbar}\Psi_{\varepsilon}(t)$  are  the examples of
the such  symmetries  [1-4]. Recently has been established  also  what
beyond that point an additional discrete transformation exists.     It
is the inversion of the speed of light $t \to t$,  $ {\bf x } \to {\bf
x}$,  $c \to -c$ or $x^0 \to x^0$,  ${\bf x} \to {\bf x}$,  $c \to -c$
where  $x^0=ct$  \cite{Kot92},  \cite{Kot95}.  Let  us  designate  the
appropriate  symmetry  by  the symbol $Q$ and name it the "minus c" or
$"c \to -c"$ symmetry.  The equation of light cone $c^2 t^2 - {\bf  x}
^2=0  \to  {(-c)}^2  t^2 - {\bf x}^2=0$ may demonstrate the example of
existing of such symmetry.  "Minus c" symmetry is also inherent in the
D'Alembert equation,  the Maxwell equations,  the equation of movement
of a charge particle in electromagnetic field,  the Schr\"odinger  and
Klein-Gordon-Fock  equations  \cite{Kot92}.  It is shown also that $"c
\to -c"$ symmetry is connected closely  with  the  charge  conjugation
$C$, which  may  be interpreted as the evidence of invariance of Dirac
equation with respect  to  the  $QPT$  composition  \cite{Kot95}.  The
purpose  of  the present work is the further study of relation between
the  transformation  of  charge  conjugation  $C$  and  $"c  \to  -c"$
inversion and possible interpretation of obtained results.

\section {Group of discrete transformations of space, time, the speed 
of light $G_8$ }
\label{s2}
\begin{sloppypar}
\par Let us introduce the 5-dimensional space of events $V^5 (x^0,{\bf
x},c)$ and  two  hyperplanes  with $c=+3.10^{10} cm/s$ and $c=-3.10^{10}
cm/s$ in it and construct four matrices of dimension 5X5:
\end{sloppypar}
\begin{equation}
\label{f1}
E=\left\lgroup\matrix { 1&0&0\cr 0&I&0\cr 0&0&1\cr}\right\rgroup;
\alpha^T=\left\lgroup\matrix { -1&0&0\cr 0&I&0\cr 0&0&1\cr}\right\rgroup;
\alpha^P=\left\lgroup\matrix { 1&0&0\cr 0&-I&0\cr 0&0&1\cr}\right\rgroup;
\alpha^Q=\left\lgroup\matrix { 1&0&0\cr 0&I&0\cr 0&0&-1\cr}\right\rgroup
\end{equation}
Here $I$  is  the  3X3  unit  matrix;   $\alpha^{T^2}$=$\alpha^{P^2}$=
$\alpha^{Q^2}$=$E$;
$[\alpha^T,\alpha^P]=[\alpha^T,\alpha^Q]=[\alpha^P,\alpha^Q]      =0$.
Matrices  and  their products form the cyclic Abelian group and induce
the 8-dimensional group of discrete transformations of time, space and
speed of light $G_8$ in the 5-space of events $V^5 (x^0,{\bf x},c)$:
\begin{equation}
\label{f2}
x^{a}{'}={(\alpha^{T^k} \alpha^{P^l} \alpha^{Q^m})}^{ab}
x^b; \ { } a,b=0,1,2,3,5
\end{equation}
The matrices $\alpha^P$, $\alpha^T$, $\alpha^Q$ are in conformity with
operators  $P$,  $T$,  $Q$  acting on field functions of the equations
studied at replacement of variables (\ref{f2}).
\par We  consider  the  place  of  group  $G_8$  and  its subgroups in
symmetry theory of classical and quantum equations.

\section {Discrete symmetry of Maxwell equations}
\label{s3}
We take  the  one-charge  Maxwell equations and shall consider them in
the 5-dimensional space $V^5 (x^0,{\bf x},c)$ on hyperplanes with $+c$
and $-c$.  The each hyperplane can be interpreted as  the  4-Minkowski
subspace  as  far  as the sign of speed of light does not influence on
the metric tensor $g_{ab}=diag(+,-,-,-)$ in view of raising to  second
power the  differentials  of  coordinates in the expression of squared
interval ${ds}^2={dx^0}^2-{d{\bf x}}^2={(-dx^0)}^2-{d{\bf  x}}^2$.  We
have on $+c$ hyperplane as follows:
\begin{equation}
\label{f3}
\begin{array}{cc}
\vspace{1mm}
\displaystyle
\nabla X {\bf H}-\partial_0{\bf E}=4\pi {\bf J}; \ \nabla.\ {\bf H}=0 \\
\displaystyle
\nabla X {\bf E}+\partial_0{\bf H}=0; \ \nabla.\ {\bf E}=4\pi \rho
\end{array}
\end{equation}
Here $x^0=ct$,  $x^{1,2,3}=x,  y,  z$,  ${\bf  E}$,  ${\bf H}$ are the
electrical and magnetic field;  $\rho$  is  the  density  of  electric
charge $e$;  ${\bf J}=\rho {\bf v}/c$ is the density of current; ${\bf
v}$ is the speed of charge;  ${\bf E}=-\partial_0 {\bf A}-\nabla\phi$,
${\bf H} =\nabla X{\bf A}$;  $A=(\phi,{\bf A})$ is  the  4-dimensional
potential.
\par The  statement  takes  place:  the   group   of   transformations
(\ref{f2}) is the group of discrete symmetry of Maxwell equations.
\begin{sloppypar}
\par The  proof  is  convenient to conduct with help of 16-dimensional
function $\Phi^{e}(x^0,{\bf x},c)=column(0,{\bf E},0,{\bf H},\rho,{\bf
j},\phi,{\bf A})$  wrote  on  the  $+c$ hyperplane and labeled by the
discrete top  index  $e$  (electric  charge).  The  Maxwell  equations
are transformed  to themselves if the function $\Phi^e$ is transformed
by the rules:
\end{sloppypar}
\begin{equation}
\label{f4}
\begin{array}{l}
\displaystyle
T_1\Phi^{ e}(x^0,{\bf x},c)=\Phi^{e}_{T1}(-x^0,{\bf x},c)=
column{(0,+{\bf E},0,-{\bf H},+\rho,-{\bf J},+\phi,-{\bf A})}_
{(-x^0,{\bf x},c)}; \\
\displaystyle
\vspace{1mm}
T_2\Phi^{e}(x^0,{\bf x},c)=\Phi^{-e}_{T2}(-x^0,{\bf x},c)=
column{(0,-{\bf E},0,+{\bf H},-\rho,+{\bf J},-\phi,+{\bf A})}_
{(-x^0,{\bf x},c)}; \\
\displaystyle
P_1\Phi^{e}(x^0,{\bf x},c)=\Phi^{e}_{P1}(x^0,-{\bf x},c)=
column{(0,-{\bf E},0,+{\bf H},+\rho,-{\bf J},+\phi,-{\bf A})}_
{(x^0,-{\bf x},c)}; \\
\displaystyle
\vspace{1mm}
P_2\Phi^{e}(x^0,{\bf x},c)=\Phi^{-e }_{P2}(x^0,-{\bf x},c)=
column{(0,+{\bf E},0,-{\bf H},-\rho,+{\bf J},-\phi,+{\bf A})}_
{(x^0,-{\bf x},c)}; \\
\displaystyle
Q_1\Phi^{e}(x^0,{\bf x},c)=\Phi^{-e}_{Q1}(x^0,{\bf x},-c)=
column{(0,-{\bf E},0,-{\bf H},-\rho,-{\bf J},-\phi,-{\bf A})}_
{(x^0,{\bf x},-c)}; \\
\displaystyle
Q_2\Phi^{e}(x^0,{\bf x},c)=\Phi^{e}_{Q2}(x^0,{\bf x},-c)=
column{(0,+{\bf E},0,+{\bf H},+\rho,+{\bf J},+\phi,+{\bf A})}_
{(x^0,{\bf x},-c)}
\end{array}
\end{equation}
The given ratios generalize the ones known in literature \cite{Ber68},
\cite{Ros73}, \cite{Str75}, \cite{Fus83}, \cite{Ber89}. In addition to
Maxwell  equations  they  keep  invariance  of D'Alembert equation for
4-potential and the  equation  of  movement  of  charged  particle  in
electromagnetic  field.  By  this  they  form the discrete symmetry in
classical electrodynamics \cite{Kot92}. Due to the ratios
\begin{equation}
\label{f5}
\begin{array}{c}
\vspace{1mm}
\displaystyle
{P_1}^2\Phi={P_2}^2\Phi={T_1}^2\Phi={T_2}^2={ Q_1}^2\Phi={Q_2}^2\Phi=E\Phi; \\
\vspace{1mm}
\displaystyle
{ P_1 P_2}\Phi= { T_1 T_2}\Phi= { Q_1 Q_2}\Phi; \\
\vspace{1mm}
\displaystyle
[ P_1 T_1, P_2 T_2]\Phi= [ P_1 Q_1, P_2 Q_2]\Phi= [ T_1 Q_1, T_2 Q_2]\Phi= \\
\displaystyle
[ P_1 T_2, P_2 T_1]\Phi= [ P_1 Q_2, P_2 Q_1]\Phi= [ T_1 Q_2, T_2 Q_1]\Phi=0
\end{array}
\end{equation}
the number of different symmetries is equal to 16 (coincides with  the
number  of  the different combinations of Dirac matrices plus the unit
matrix in quantum theory). It is the symmetries
\begin{equation}
\label{f6}
\begin{array}{c}
\displaystyle
E, \ P_1, \ P_2, \ T_1, \ T_2, \ Q_1, \ Q_2, \ P_1 T_1, \ P_1 T_2, \ P_1 Q_1,
\ P_1 Q_2, \\
\displaystyle
T_1 Q_1, \ T_1 Q_2, \ Q_1 Q_2, \ P_1 T_1 Q_1, \ P_1 T_1 Q_2
\end{array}
\end{equation}
All other combinations, total  number of which equals  $N=1 + { C_6}^1
+ { C_6}^2 + { C_6}^3 + { C_6}^4 + { C_6}^5 + { C_6}^6  =64$,  can  be
expressed   through  combinations (\ref{f6})   for  example  $P_1  Q_1
Q_2\Phi=P_2 Q_2 Q_2\Phi=P_2\Phi$,  $P_1 P_2 T_1 T_2\Phi= E\Phi$,  $P_1
P_2 T_1 T_2 Q_1 Q_2\Phi=Q_1 Q_2\Phi$ and etc...  The symmetries of the
type $E$,  \ $P_1$, \ $T_1$, \ $C$, \ $CT_1$, \ $CT_2$, \ $P_1 T_1$, \
$CP_1  T_1$ were studied in \cite{Sch57},  \cite{Ber68}, \cite{Shi72},
for example, in connection with physics problems. Additional $P_2$ and
$T_2$ symmetries  were  investigated  in  \cite{Ros73},  \cite{Str75},
\cite{Ber89}.  Symmetry  of the $Q$ type and combinations $QT$,  $QP$,
$QPT$ were discussed in  the  works  \cite{Kot92},  \cite{Kot95}.  The
restriction on the number of symmetries are usually connected with the
requirement of scale property of electric charge with respect to space
inversion and time reversal, which in symmetry approach is dispensable
\cite{Ros73}, \cite{Str75}, \cite{Ber89}.
\par  Below we shall study the symmetries connected with the inversion 
of the speed of light.

\section {The free Maxwell equations}
\label{s4}
The inversion of the speed of light is the particular case of discrete
transformations  of group $G_8$ and thus forms the symmetry of Maxwell
equations.  Let us  consider  the  properties  of  charge  conjugation
induced  by  the  $c  \to  -c$  inversion  in case of the free Maxwell
equations.

\subsection {The charge conjugation in classical sense}
\label{ss4.1}
We turn attention to the combination $Q_1 Q_2$ which we designate by 
symbol $C_e$ and which induces the transformations
\begin{equation}
\label{f7}
C_e\Phi^{e}(x^0,{\bf x},c)=\Phi^{ -e}(x^0,{\bf x},c)=
column{(0,-{\bf E},0,-{\bf H},-{\bf j},-\rho,-\phi,-{\bf A})}_
{(x^0,{\bf x},c)}
\end{equation}
Operator  $C_e$  change  the sign of electrical charge and may be
interpreted as the operator of charge conjugation type.
In the  case  of  the  free  fields  the  charge  conjugated  function
$\Phi_{Ce}=column(0,-{\bf E},0,-{\bf H},0,0,0,-\phi,-{\bf A})_
{(x^0,{\bf x},c)}$ is characterized by the change of the signs of  the
electric  and magnetic fields ${\bf E}={\bf l} \ exp [-i(k^0 x^0 -{\bf
k}{\bf x})]$ and ${\bf H}={\bf m} \ exp [-i( k^0 x^0-{\bf k}{\bf x})]$
where ${\bf m}={\bf n}$x${\bf l}$. By means of replacement $k^0=\omega
/c={\cal E}/\hbar c=p^0/\hbar$,  ${\bf  k}=k^0{\bf  n}={\bf  p}/\hbar$
($\omega$  is  the frequency of electromagnetic field,  $\hbar$ is the
Plank constant,  ${\bf n}$ is the guiding wave vector,  ${\cal E}$  is
the energy of field, ${\bf p}$ is the momentum of field) the result of
charge conjugation (\ref{f7}) may be written as follows
\begin{equation}
\label{f9}
\begin{array}{c}
\displaystyle
C_e\Psi_{ p \ {\bf n} \ {\bf l}} (x^0,{\bf x},c)=C_e column (0,
{\bf E},0,{\bf H})=column(0,-{\bf E},0,-{\bf H})= \\
column(0,-l_1,-l_2,-l_3,0,-m_1,-m_2,-m_3) \ e^{-\frac{i}{\hbar}
(p^0 x^0 - {\bf p} {\bf x})}=\Psi_{ p \ {\bf n} \ -{\bf l}},
\end{array}
\end{equation}
Here $\Psi \in \Phi$.  The charge conjugation $C_e$ changes the  signs
of  polarization  vectors ${\bf l}$ and ${\bf m}$ and keeps the signs
of the energy $C_e{\cal E}={\cal E}$,  momentum $C_e{\bf  p}={\bf  p}$
and guiding wave vector of  field  $C_e{\bf  n}={\bf  n}$.  It  is  in
agreement  with  the  behaviour of the density of the field energy and
the field momentum calculated directly from Maxwell equations  $W=(E^2
+  H^2)/8\pi$,  ${\bf S}=c({\bf E}X{\bf H})/4\pi$.  By this the charge
conjugation $C_e=Q_1 Q_2$,  being the transformation  of  symmetry  of
Maxwell  equations,  does not result in negative energies.  It differs
from the  charge  conjugation  $C$  in  quantum  theory  \cite{Sch57},
\cite{Ber68}.  By given attribute it is possible to identify the $C_e$
conjugation as the charge conjugation  in  the  classical  sense.  The
result of the $C_e$ conjugation coincides with known one \cite{Ber68}.

\subsection {The charge conjugation in quantum sense}
\label{ss4.2}
Let us write the Maxwell equations in form of the Dirac equation  with
help of the 8-dimensional function $\Psi=column(0,  E_1,  E_2, E_3, 0,
H_1, H_2, H_3)_ {(x^0,{\bf x},c)}$ \cite{Fus83}:
\begin{equation}
\label{f10}
\gamma^a p_a\Psi(x^0,{\bf x},c)=(i\hbar\gamma^0\partial_0 + i\hbar\gamma.
\nabla)\Psi(x^0,{\bf x},c) =0
\end{equation}
Here $x^a=(ct,x,y,z); \ g_{ab}=diag(+,-,-,-); \ p_a=i\hbar\partial/
\partial x^a; \ a,b=0,k; \ k=1,2,3$; \ the summation is carried out over
the twice repeating index; \ $\gamma^a=(\gamma^0,\gamma);                                         \
\gamma\equiv\gamma^k=(\gamma^1,\gamma^2,\gamma^3)$ are the  8-matrices
\cite{Fus83}:
\begin{equation}
\label{f11}
\gamma ^0=\left\lgroup\matrix{0&I\cr I&0\cr}\right\rgroup;
\gamma ^k=\left\lgroup\matrix{\alpha^k&0\cr 0&-\alpha^k\cr}\right\rgroup; 
\gamma ^5=\left\lgroup\matrix{0&-I\cr -I&0\cr}\right\rgroup
\end{equation}
\begin{equation}
\label{f12}
\alpha^1=
\left\lgroup\matrix{0&1&0&0\cr-1&0&0&0\cr0&0&0&-1\cr0&0&1&0\cr}\right\rgroup;
\alpha^2=
\left\lgroup\matrix{0&0&1&0\cr0&0&0&1\cr-1&0&0&0\cr0&-1&0&0\cr}\right\rgroup;
\alpha^3=
\left\lgroup\matrix{0&0&0&1\cr0&0&-1&0\cr0&1&0&0\cr-1&0&0&0\cr}\right\rgroup
\end{equation}
($I$ -the unit 4-matrix). The gamma matrices have the following properties:
\begin{equation}
\label{f13}
\begin{array}{c}
\displaystyle
\vspace{1mm}
\gamma ^a \gamma ^b + \gamma ^b \gamma ^a = 2g^{ab}; \ {}
\gamma ^a \gamma ^5 + \gamma ^5 \gamma ^a = -2g^{a0}; \\
\displaystyle
(\gamma ^0)^+=\gamma ^0; \ {} (\gamma ^{1,2,3})^+=-(\gamma ^{1,2,3}); \ {}
(\gamma ^0)^2=1; \ {} (\gamma ^{1,2,3})^2=-1; \\
\displaystyle
(\gamma^{0,1,2,3})^*=\gamma^{0,1,2,3}; \ {} 
(\gamma^{0})^T=\gamma^{0};  \ {}
(\gamma^{1,2,3})^T=-\gamma^{1,2,3}; \\
\displaystyle
\gamma ^5=\gamma^0 \gamma ^1 \gamma ^2 \gamma ^3; \ {}  (\gamma ^5)^+=
\gamma^5; \ {} (\gamma^5)^*=\gamma^5; \ {}   (\gamma ^5)^2=1
\end{array}
\end{equation}
Meaning the conversion of Maxwell equations into  Dirac  equation,  we
use  the  operator  of  charge conjugation $C$ from quantum theory and
define the action of the operator $C$ by analogy with \cite{Sch57}:
\begin{equation}
\label{f14}
C\Psi(x^0,{\bf x},c)=U_C {\overline\Psi}^T (x^0,{\bf x},c)
=U_C\gamma^0 \Psi^ * (x^0,{\bf x},c)
\end{equation}
Here $U_C$ is the matrix of charge conjugation,  $\overline\Psi=\Psi^+
\gamma^0$  is  the  Dirac  conjugated  function,  $*$  is  the complex
conjugation,  $\tt T$ is the transposition.  
\par We  take  the  equation  (\ref{f10}) and rewrite it for the Dirac
conjugated  function  $\overline\Psi$.   Making   the   transposition,
multiplying the result at the left by the matrix $U_C$ and  using  the
property of the gamma-matrices (\ref{f13}) we find:
\begin{equation}
\label{f15}
\begin{array}{c}
\displaystyle
\gamma^ap_a\Psi=0 \to \overline\Psi\gamma^ap_a=0 \to (i\hbar\gamma^{0T}
\partial_0 + i\hbar\gamma^T .\nabla ){(\Psi^+ \gamma^0)}^T
\to \\ (i\hbar U_C\gamma^{0T} {U_C}^{ -1}\partial_0 +
i\hbar U_C\gamma^T {U_C}^{-1} .\nabla)U_C {(\Psi^+ \gamma^0)}^T =0
\end{array}
\end{equation}
It is  seen   that  equation  (\ref{f15})  for  the  charge conjugated
function  $U_C{\overline\Psi}^T$ coincides with the initial   equation
(\ref{f10}) if the matrix $U_C$ satisfies the conditions:
\begin{equation}
\label{f16}
U_C\gamma^{aT} {U_C}^{-1}=\gamma^a \to
U_C\gamma^{0} {U_C}^{-1}=\gamma^0; \
U_C\gamma^{k} {U_C}^{-1}=-\gamma^k; \ k=1,2,3
\end{equation}
As far  as  $U_C\gamma^k  +  \gamma^k  U_C=0$   in   accordance   with
(\ref{f13})   we can put $U_C=\lambda\gamma^0$, where $\lambda$ is the
factor of proportionality.  The expression for the  charge  conjugated
function takes the form:
\begin{equation}
\label{f17}
\Psi_C=\lambda\gamma^0 {(\Psi^+ \gamma^0)}^T =\lambda\gamma^0\gamma^0
\ \Psi^ * =\lambda\Psi^* =\lambda \ column (0, {E_1}^*, {E_2}^*, {E_3}^*,
0, {H_1}^*, {H_2}^*, {H_3}^*)
\end{equation}
The formula  (\ref{f17})  coincides  with  the  result \cite{Fus83} if
$\lambda=1$.  Further we write the  function  describing  the  initial
photon state by analogy with \cite{Sch57}, \cite{Ber68}
\begin{equation}
\label{f18}
\displaystyle
\Psi_{p \ {\bf n}\ {\bf l}}=\frac {1}{\sqrt 2} column(0, l_1, l_2,
l_3, 0, m_1, m_2, m_3 ) \ e^{-\frac {i}{\hbar} (p^0 x^0 - {\bf p}{\bf x})},
\end{equation}
where $\Psi^+ \Psi=1$,    $p=({\cal E}/c,{\cal E}{\bf n}/c)$    is the
4-momentum, ${\cal E}=\hbar\omega$,  \ ${\bf p}={\cal E}{\bf n}/c$ are
the energy and momentum of photon,  ${\bf n}$ is the guiding vector of
photon,  ${\bf l}$ is the vector of electrical polarization. Acting on
function (\ref{f18}) by the operator $C$,  we find the expression  for
charge conjugated function in agreement with formula (\ref{f17}):
\begin{equation}
\label{f19}
C\Psi_{p \ {\bf n} \ {\bf l}} = \Psi_{-p \ {\bf n} \ \lambda {\bf l}} =
\frac {\lambda}{\sqrt 2} column( 0, l_1, l_2, l_3, 0, m_1, m_2, m_3 ) \
e^{\frac {i}{\hbar}(p^0 x^0 - {\bf p}{\bf x})},
\end{equation}
The $\lambda$ value is equal to $(\pm 1,\pm i)$ as it follows from the
condition  of  the  $\Psi$  function  normalization.  Similarly to the
solution of Dirac equation for  a  particle  with  nonzero  rest  mass
\cite{Sch57}, \cite{Ber68}, \cite{Shi72}, the photon charge conjugated
function is possible to be considered as  the  function  describing  a
particle  with  negative  energy  ${\cal E}=-\hbar\omega$ and opposite
momentum   ${\bf   p}=-(\hbar\omega/c){\bf   n}$.   With   the   field
interpretation of the energy and momentum $\lambda$ is equal $-i$, for
example. In this case the field densities of energy and field momentum
are $-(E^{*2} + H^{*2})/8\pi,  \ -c(E^* XH^* )/4\pi$ i.e. are negative
as in quantum theory.
\par We introduce the designation $\Psi_C=\Psi_{-p \ {\bf n} \ \lambda
{\bf l}}$ and in spirit of the Dirac interpretation  of  the  solution
with  negative  energy shall consider the $\Psi_C$ solution as the one
describing directly not observable vacuum photon.  It follows from the
formula  (\ref{f19})  that  besides  negative  energy and momentum the
vacuum photon is characterized by initial sign of  guiding  vector  of
propagation  ${\bf  n}$  and by the vectors of electrical and magnetic
field polarization ${\bf l}$ and ${\bf m}$ depending from  the  choice
of  the  $\lambda$  value.  It  follows  also  from expressions of the
quantum      8-currents      $j^a=\overline\Psi\gamma^a\Psi$       and
${j^a}_C=\overline\Psi_C\gamma^a\Psi_'$,     that     the    relations
$j^0={j^0}_C=1$,  $j^k ={j^k}_C=(n^1,  \ n^2,  \ n^3)$ take  place  in
agreement  with the property of absence of photon charge \cite{Ber68}.
The action of the  charge  conjugation  operator  $C$  on  the  charge
conjugated  function  $\Psi_C$  transforms  the latter to the function
describing  the  state   with   positive   energy   $C\Psi_C=i\gamma^0
{({\Psi_C}^+ \gamma^0)}^T= i{\Psi_C}^* =\Psi_{p \ {\bf n} \ {\bf l}}$.
This state may be interpreted as the  antiphoton  identical  with  the
photon as result of its neutrality.
\par Further  we shall find how the operator of charge conjugation $C$
may be connected with the operator of conjugation $Q$ induced  by  the
inversion  of  the  speed of light $c \to -c$.
\par Let us define the conjugation $Q$ as follows:
\begin{equation}
\label{f20}
Q\Psi(x^0,{\bf x},c)=U_Q{\overline\Psi}^T (x^0,{\bf x},-c)=
U_Q\gamma^0 \Psi^* (x^0,{\bf x},-c), \ \hbar \to  -\hbar
\end{equation}
The inversion of the speed of light and Plank constant do  not  change
the  equation  (\ref{f10})  because  of  absence  of photon rest mass.
Consequently the  conjugation  $Q$  transforms  formally  the equation
(\ref{f10}) into itself by means of the same matrix as in the case  of
the charge conjugation $U_Q=U_C= \lambda\gamma^0$.  The law of initial
photon function $\Psi_{p \ {\bf n} \ {\bf l}}$ transformation is:
\begin{equation}
\label{f21}
\begin{array}{c}
\displaystyle
\vspace{1mm}
Q\Psi_{p \ {\bf n} \ {\bf l}}=\Psi_{-p \ {\bf n} \ \lambda {\bf l}}=
\frac{\lambda}{\sqrt 2} column(0, l_1, l_2, l_3, 0, m_1, m_2, m_3)  \
{[e^{-\frac{i}{-\hbar} (-p^0 x^0 + {\bf p}{\bf x})}]}^* = \\
\displaystyle
\frac{1}{\sqrt 2} column(0,-l_1,-l_2,-l_3, 0,-m_1,-m_2,-m_3)  \
{e^{\frac{i}{\hbar} (p^0 x^0 - {\bf p}{\bf x} + \hbar\pi/2)}}
\end{array}
\end{equation}
Here we put  $\lambda=-i$  and  take  that  with  inversion  of  Plank
constant    all    components   of   4-momentum   change   the   signs
$i\hbar\partial_a \to -i\hbar\partial_a$ owing  to  constancy  of  the
space  variables $x^a$.  The obtained result is useful to compare with
the result described by formula (\ref{f9})  for  the  case  of  charge
conjugation  in  the  classical  sense.  The  distinction  consists in
occurrence  in  the  formula  (\ref{f21})  the  normalization   factor
$1/\sqrt 2$  and sign reversing and displacement of the phase.
Besides we should also note that  the  formula  (\ref{f21})  coincides
with the formula (\ref{f19}) owing to the following interrelation:
\begin{equation}
\label{f22} C\Psi_{p \ {\bf n} \ {\bf l}} (x^0,{\bf x},c)=
Q\Psi_{p \ {\bf n}  \  {\bf l}} (x^0,{\bf x},c)
\end{equation}
The important  conclusion  may  be  formed  from  here.   The   charge
conjugated  function  describing the vacuum photon state with negative
energy on $+c$ hyperplane coincides with the function  describing  the
free photon state with positive energy on the $-c$ hyperplane.
\begin{equation}
\label{f23}
\Psi_C (x^0,{\bf x},c)=\Psi_Q (x^0,{\bf x},-c) \to
\Psi_{-p \ -{\cal E} \ {\bf n} \ \lambda {\bf l}} (x^0,{\bf x},c)=
\Psi_{-p \ +{\cal E} \ {\bf n} \ \lambda {\bf l}} (x^0,{\bf x},-c)
\end{equation}
One can see that  the  vacuum  photon  from  the $+c$ hyperplane    is
equivalent to the free photon from the $-c$ hyperplane. It is possible
to admit that this is the same object between characteristics of which
exist the thin distinctions which depends on its interpretation.
\par In the case of the vacuum  interpretation  we  may  believe  that
photon  is located on $+c$ hyperplane with the positive Plank constant
$\hbar>0$.  Its energy is negative  ${\cal  E}=-\hbar\omega  <0$,  the
frequency is negative $\omega <0$,  the 4-momentum components have the
opposite signs $p^a=(-p^0,-{\bf p})$.  With  the  field  approach  the
4-dimensional   wave   vector   components  have  the  opposite  signs
$k^a=(-\omega /c,-\omega {\bf n})/c$.  The photon is in  condition  of
vacuum  movement  with  the  positive  speed of light and the negative
energy.
\par In  the  case of the "minus-c" interpretation we may believe that
photon is located on the  $-c$  hyperplane  with  the  negative  Plank
constant    $\hbar    <0$.    Its    energy    is    positive   ${\cal
E}=(-\hbar)(-\omega)>0$,  the frequency is negative $\omega  <0$,  the
4-momentum  components  have the opposite signs $p^a=(-p^0,-{\bf p})$.
With the field approach the 4-dimensional wave vector components  have
the  initial  signs $k^a=(-\omega /(-c),-\omega {\bf n} /(-c))=(\omega
/c,  \ \omega {\bf n} /c)$.  The photon is in condition  of  the  free
movement with the negative speed of light and the positive energy.
\par Both interpretation are in agreement with the  ratio  (\ref{f23})
which describes the interrelation between the $C$ conjugated state and
the $Q$ conjugated state of photon.  Their existence are reflected the
fact  that  the  charge conjugation in the sense of quantum theory for
the  case  of  Maxwell  equations  can  be  interpreted  also  as  the
consequence  of  the  invariance  of  these  equations with respect to
inversion of the speed of light $c \to -c$.  As we can see below,  the
similar property takes place not only for photon but also for electron
states from the Dirac equation \cite{Kot95}.  We note  only  that  the
operation  of  the  $Q$  conjugation  in our case differs from the $Q$
conjugation \cite{Kot95} where the Plank constant kept  the  invariant
significance.  The  choice  the $Q$ conjugation in form of (\ref{f14})
seems to be more correct for the two reasons.
\begin{itemize}
\item In  the case of the Plank constant invariance the thin structure
constant $\alpha =e^2/\hbar c$ is not invariant as far as $Q(e^2/\hbar
c)=-e^2/\hbar c$ that is undesirable.
\item In the classical electrodynamics with the non-invariant speed of
light the new invariants take place:  $\hbar c$ \cite{Kot70} or $\hbar
/ c$ \cite{Hsu76}.
\end{itemize}
Below we  select  the variant \cite{Kot70} according to which the rest
mass of particle $m$ is transformed as $Q(m)=m'=mc^2/{(-c)}^2=m$, that
is, keeps the invariant significance on the hyperplane $-c$.

\section{The Dirac equation}
\label{s5}
Let us introduce the Dirac equation in the form \cite{Sch57}, \cite{Ber68}
\begin{equation}
\label{g2}
(\gamma ^a p_a -mc)\psi (x^0,{\bf x},c) = (i\hbar \gamma ^0 \partial _0 +
i\hbar \gamma .\nabla -mc)\psi (x^0,{\bf x},c) =0
\end{equation}
\noindent Here $x^a=(ct,x,y,z)$;  $g_{ab}=diag(+,-,-,-)$;  $\gamma ^a=
(\gamma  ^0,\gamma)$;  $\gamma=(\gamma^1,\gamma ^2,\gamma ^3)$ are the
Dirac matrices;  $p_a=i\hbar \partial  /\partial  x^a=i\hbar  \partial
_a$;  $a=0,1,2,3$;  the  summation  is  carried  out  over  the  twice
repeating index; $\Psi =column(\phi,\chi)$; $\phi =column(\phi _1,\phi
_2)$; $\chi =column(\chi _1,\chi _2)$. The gamma matrices
\begin{equation}
\label{g4}
\gamma ^0=\left\lgroup\matrix{I&0\cr 0&-I\cr}\right\rgroup;
\gamma ^{1,2,3}=\left\lgroup\matrix{0&\sigma _{x,y,z}\cr -\sigma
_{x,y,z}&0\cr}\right\rgroup;
\gamma ^5=\left\lgroup\matrix{0&-I\cr -I&0\cr}\right\rgroup
\end{equation}
\begin{equation}
\label{g5}
\sigma _x=\left\lgroup\matrix{0&1\cr1&0\cr}\right\rgroup;
\sigma _y=\left\lgroup\matrix{0&-i\cr i&0\cr}\right\rgroup;
\sigma _z=\left\lgroup\matrix{1&0\cr0&-1\cr}\right\rgroup
\end{equation}
where I  is  the  unit  two dimensional matrix,  satisfy the relations
\cite{Sch57}, \cite{Ber68}:
\begin{equation}
\label{g3}
\begin{array}{c}
\displaystyle
\vspace{1mm}
\gamma ^a \gamma ^b + \gamma ^b \gamma ^a = 2g^{ab}; \ {}
\gamma ^a \gamma ^5 + \gamma ^5 \gamma ^a =0; \\
\displaystyle
(\gamma ^0)^+=\gamma ^0; \ {} (\gamma ^{1,2,3})^+=-(\gamma ^{1,2,3}); \ {}
(\gamma ^0)^2=1; \ {} (\gamma ^{1,2,3})^2=-1; \\
\displaystyle
(\gamma^{0,1,3})^*=\gamma^{0,1,3}; \ {} (\gamma^2)^*=-\gamma^2; \ {}
(\gamma^{0,2})^T=\gamma^{0,2};  \ {}
(\gamma^{1,3})^T=-\gamma^{1,3}; \\
\displaystyle
\gamma ^5=-i\gamma^0 \gamma ^1 \gamma ^2 \gamma ^3; \ {}  (\gamma ^5)^+=
\gamma^5; \ {} (\gamma^5)^*=\gamma^5; \ {}   (\gamma ^5)^2=1
\end{array}
\end{equation}
\noindent In accordance with \cite{Sch57},  \cite{Ber68} and  formulas
(\ref{f14}), (\ref{f20}) we define the action of operators $C$ and $Q$
on the solution $\psi $ in the form:
\begin{equation}
\label{g8}
C\psi (x^0,{\bf x},c)=U_C {\overline \psi}^T (x^0,{\bf x},c)=
U_C \gamma ^0 \psi ^* (x^0,{\bf x},c);
\end{equation}
\begin{equation}
\label{g9}
Q\psi (x^0,{\bf x},c)=U_Q {\overline \psi}^T (x^0,{\bf x},-c)=
U_Q \gamma ^0 \psi ^* (x^0,{\bf x},-c); \hbar \to -\hbar
\end{equation}
\noindent Here  $U_C$  and $U_Q$ are the corresponding matrices of the
transformations of the bispinor  ${\overline \psi}^T (x^0,{\bf  x},c)$
and ${\overline\psi}^T (x^0,{\bf x},-c)$;    ${\overline \psi} =\psi^+
\gamma^0$;  ${\tt  T}$  is  the  transposition;  $*$  is  the  complex
conjugation.  We  take  Eq.  (\ref{g2})  and perform the $Q$-inversion
(\ref{g9}) in it:
\begin{equation}
\label{g10}
\begin{array}{c}
\displaystyle
Q: \ (\gamma ^a p_a - mc)\psi (x^0,{\bf x},c)=0 \to
{\overline\psi}(x^0,{\bf x},-c)(\gamma^a p_a + mc)=0 \to \\
\displaystyle
(\gamma^{aT}p_a + mc){\overline\psi}^T(x^0,{\bf x},-c) \to
(U_Q\gamma ^{aT}U_Q ^{-1} p_a + mc)U_Q {\overline\psi}^T (x^0,{\bf x},-c)=0
\end{array}
\end{equation}
Let Eq.  (\ref{g10}) coincides with the initial Eq. (\ref{g2}) for the
transformed function $\psi _Q=U_Q{\overline\psi}^T$.  Matrix $U_Q$ has
the following property then:
\begin{equation}
\label{g11}
U_Q\gamma^{aT}U_Q^{-1}=-\gamma ^a \to U_Q\gamma^{0,2}=-\gamma^{0,2}U_Q, \
U_Q\gamma^{1,3}=\gamma^{1,3}U_Q
\end{equation}
The properties of the $U_Q$ -matrix are the same as the $U_C$  -matrix
of  the  charge conjugation $C$ \cite{Sch57},  \cite{Ber68} and we can
write:
\begin{equation}
\label{g14}
U_Q=U_C=-\gamma ^0\gamma ^2
\end{equation}
We take  into  account  this  relationship  and  make  a  table of the
transformations of the function $\psi(x^0,{\bf x},c)$ relative to  the
$Q$ -inversion and the charge conjugation $C$.
\begin{equation}
\label{g15}
\begin{array}{ll}
Berestetski, Lifshits, Pitaevski                                  &
The \ {present} \ {work}                                          \\
\vspace{3mm}
C, P, T \ {symmetries}, \ {c \to +c}, \ {\hbar \to +\hbar}, \ {\cite{Ber68}}: &
P, T, Q \ {symmetries}, \ {c \to \pm c}, \ {\hbar \to \pm\hbar}:  \\
P \psi = i\gamma ^0\psi (x^0,-{\bf x},c);                         &
P \psi = i\gamma ^0\psi (x^0,-{\bf x},c);                         \\
T \psi = -i\gamma ^1\gamma ^3\psi ^* (-x^0,{\bf x},c);            &
T \psi = -i\gamma ^1\gamma ^3\psi ^* (-x^0,{\bf x},c);            \\
\vspace{3mm}
P T \psi = \gamma ^0\gamma ^1\gamma ^3\psi ^* (-x^0,-{\bf x},c);  &
P T \psi = \gamma ^0\gamma ^1\gamma ^3\psi ^* (-x^0,-{\bf x},c);  \\
C P T \psi = i\gamma ^5 \psi (-x^0,-{\bf x},c);                   &
Q P T \psi = i\gamma ^5 \psi (-x^0,-{\bf x},-c);                  \\
C T \psi = i\gamma ^1 \gamma ^2 \gamma ^3 \psi (-x^0,{\bf x},c);  &
Q T \psi = i\gamma ^1 \gamma ^2 \gamma ^3 \psi (-x^0,{\bf x},-c); \\
C P \psi = i\gamma ^0 \gamma ^2 \psi ^* (x^0,-{\bf x},c);         &
Q P \psi = i\gamma ^0 \gamma ^2 \psi ^* (x^0,-{\bf x},-c);        \\
C \psi = \gamma ^2 \psi ^* (x^0,{\bf x},c);                       &
Q \psi = \gamma ^2 \psi ^* (x^0,{\bf x},-c)
\end{array}
\end{equation}
One can  see  that  the  charge  conjugation  $C$  corresponds  to the
$Q$-transformation so that
\begin{equation}
\label{g16}
[C,\ Q]\psi (x^0,{\bf x},c)=0
\end{equation}
Similarly, the  operations  $CPT$,  \  $CT$ and $CP$ correspond to the
$QPT$,  \ $QT$, \ $QP$ combinations respectively.
\par Let  us  consider  in  more   detail   the   mechanism   of   the
$C\leftrightarrow  Q$  correspondence.  Following  to  \cite{Ber68} we
rewrite the function  $\psi$  in  explicit  form  for  our  case  when
$c\neq1$, $\hbar\neq1$:
\begin{equation}
\label{g17}
\psi_{p\sigma}={1\over \sqrt {2p^0}}u_{p\sigma}e^{-{i\over \hbar}p.x}; \
{}
\psi_{-p-\sigma}={1\over \sqrt {2p^0}}u_{-p-\sigma}e^{{i\over\hbar}p.x};
\end{equation}
\begin{equation}
\label{g18}
u_{p\sigma}=\left\lgroup\matrix{\sqrt {p^0+mc} \ {}w\cr \sqrt {p^0-mc}{}
({\bf n}\sigma)w\cr}\right\rgroup;
u_{-p-\sigma}=\left\lgroup\matrix{\sqrt {p^0-mc}({\bf n}\sigma)w'\cr
\sqrt {p^0+mc} \ {}w'\cr}\right\rgroup
\end{equation}
Here $p^0={\cal E}/c>0$,       ${\cal E}=c\sqrt{{\bf  p}^2+m^2c^2}>0$,
$p.x=p^0x^0-{\bf p}{\bf x}$, ${\bf n}={\bf p}/p$, $w^+ w=1$, $w= ({\bf
n}\sigma )w'$,  $\overline {u}_p u_p=2mc$, $\overline {u}_{-p} u_{-p}=
-2mc$,  $c>0$.  The result of operators $C$ and $Q$ application to the
function $\psi_{p\sigma}$ is:
\begin{equation}
\label{g19}
\begin{array}{c}
\displaystyle
\vspace{2mm}
C\psi_{p\sigma {\cal E}} (x^0,{\bf x},c)=
\gamma^2 {\psi_{p\sigma {\cal E}}^* (x^0,{\bf x},c)}_
{(c \to c \ \hbar \to \hbar \ \sigma \to \sigma)} = \\
\displaystyle
\vspace{2mm}
\frac{1}{\sqrt {2p^0}}
\left\lgroup\matrix{0&\sigma_y\cr-\sigma_y&0\cr}\right\rgroup
\left\lgroup\matrix{\sqrt {p^0+mc} \ {w}^*\cr
\sqrt {p^0-mc} \ ({\bf n}{\sigma}^*){w}^*\cr}\right\rgroup 
{(e^{-\frac{i}{\hbar} ({p^0}{x^0}-{\bf p}{\bf x})})}^*= \\
\displaystyle
\vspace{2mm}
\frac{1}{\sqrt {2p^0}}
\left\lgroup\matrix{\sqrt {p^0-mc} \ ({\bf n}\sigma)w'\cr
\sqrt {p^0+mc} \ w'\cr}\right\rgroup
e^{\frac{i}{\hbar} ({p^0}{x^0}-{\bf p}{\bf x})}=
\frac{u_{-p-\sigma}}{\sqrt {2p^0}}
e^{\frac{i}{\hbar} ({p^0}{x^0}-{\bf p}{\bf x})}=
\psi_ {-p-\sigma -{\cal E}} (x^0,{\bf x},c)
\end{array}
\end{equation}
\begin{equation}
\label{gg19}
\begin{array}{c}
\displaystyle
\vspace{2mm}
Q\psi_ {p\sigma {\cal E}} (x^0,{\bf x},c)=
\gamma^2 \psi_{p\sigma {\cal E}}^* (x^0,{\bf x},c)_
{(c \to -c \ \hbar \to -\hbar \ \sigma \to -\sigma)} = \\
\displaystyle
\vspace{2mm}
\frac{1}{i\sqrt {2p^0}}
\left\lgroup\matrix{0&-\sigma_y\cr\sigma_y&0\cr}\right\rgroup
\left\lgroup\matrix{-i\sqrt {p^0+mc} \ {w}^*\cr
-i\sqrt {p^0-mc} \ ({\bf n}{\sigma}^*){w}^*\cr}\right\rgroup 
{(e^{-\frac{i}{-\hbar} ({-p^0}{x^0}+{\bf p}{\bf x})})}^*= \\
\displaystyle
\vspace{2mm}
\frac{1}{\sqrt {2p^0}}
\left\lgroup\matrix{\sqrt {p^0-mc} \ ({\bf n}\sigma)w'\cr
\sqrt {p^0+mc} \ w'\cr}\right\rgroup
e^{\frac{i}{\hbar} ({p^0}{x^0}-{\bf p}{\bf x})}=
\frac{u_{-p-\sigma}}{\sqrt {2p^0}}
e^{\frac{i}{\hbar} ({p^0}{x^0}-{\bf p}{\bf x})}=
\psi_ {-p-\sigma+{\cal E}} (x^0,{\bf x},-c)
\end{array}
\end{equation}
We take into account in the  first  expression  that  $\sigma_y  ({\bf
n}\sigma^*)=-({\bf  n}\sigma)\sigma_y$,    $-\sigma_y  w^*=w'$,  $w'^+
w'={(\sigma_y w^*)}^+ (\sigma_y w^*)=w^T{\sigma_y}^+\sigma_y w^*={(w^+
w)}^*=1$  and  in  addition  to  this  in  the  second expression that
$p^0={\cal E}/(-c)<0$,             ${\bf p}={\cal E}(-{\bf v})/c^2<0$.
Analogously  to  result  (\ref{f22})  for  photon  states  we have for
electron ones
\begin{equation}
\label{g20}
C\psi_{p\sigma {\cal E}}(x^0,{\bf x},c)=Q\psi_{p\sigma \cal E}(x^0,{\bf x},c)
\to
\psi_{-p-\sigma -{\cal E}}(x^0,{\bf x},c)=\psi_{-p-\sigma +{\cal E}}
(x^0,{\bf x},-c)
\end{equation}
The difference  consists  in  that the electromagnetic field describes
the neutral  particles  (photons)  and  the  $\psi_{p\sigma {\cal E}}$
field describes  the  charge  particles  (electrons,  positrons).   We
consider this case in more detail.

\subsection{The Dirac equation for a charged particle}
\label{ss5}
\par Let  us  introduce  the Dirac equation for a charge particle with
spin 1/2 in electromagnetic field:
\begin{equation}
\label{g21}
(\gamma ^a p_a - mc)\psi (x,c)=(e/c)\gamma ^a A_a \psi (x,c)
\end{equation}
\noindent where  $x=(x^0,{\bf  x})$,  $e$ is the charge of a particle,
$A^a=(A^0,{\bf A})$ is the 4-potential,  $\gamma$  -are  the  matrices
(\ref{g4}).  Let  us  subject  the  equation  (\ref{g21})  to  the $Q$
transformation  taking   into   account   the   formulas   (\ref{g9}),
(\ref{g14})  on  the assumption that an electrical charge is a scalar;
the vector - potential is a polar vector relative to the  replacements
both  ${\bf x}  \to  -{\bf x}$,  and  $t \to -t$,  and  $c \to -c$
\cite{Kot95}.
\begin{equation}
\label{g22}
\begin{array}{c}
\displaystyle
Q(x^0,{\bf x},c,e,A^0,{\bf A})=(x^0,{\bf x},-c,e,A^0,{\bf A}); \\
\displaystyle
\vspace{1mm}
Q\psi (x^0,{\bf x},c)=\psi _Q (x^0,{\bf x},-c)=U_Q{\overline\Psi}^T
(x^0,{\bf x},-c)
\end{array}
\end{equation}
Taking the  Dirac  conjugate  equation,  making   the   transposition,
multiplying  on  the matrix $U_Q$ at the left and using the properties
of the $\gamma$ -matrices (\ref{g3}) we have
\begin{equation}
\label{g23}
\begin{array}{c}
\displaystyle
Q: \ (\gamma ^a p_a - mc)\psi  = (e/c)\gamma ^a A_a \psi \to \\
\displaystyle
{\overline\psi}(x^0,{\bf x},-c)(\gamma^a p_a +mc)=-(e/c){\overline\psi}
(x^0,{\bf x},-c)(A^0 \gamma^0 - {\bf A}.\gamma) \to \\
\displaystyle
(i\hbar{\gamma^0}^T + i\hbar{\gamma}^T.\nabla + mc){\overline\psi}^T
(x^0,{\bf x},-c)=-(e/c)({\gamma^0}^T A^0 - {\gamma}^T {\bf A}) \to \\
\displaystyle
(i\hbar U_Q{\gamma^0}^T {U_Q}^{-1} + i\hbar U_Q {\gamma}^T {U_Q}^{-1}.\nabla
+ mc) U_Q {\overline\psi}^T (x^0,{\bf x},-c)= \\
\displaystyle
-(e/c)(U_Q {\gamma^0}^T {U_Q}^{-1} A^0 - U_Q {\gamma}^T {U_Q}^{-1} {\bf A})
U_Q {\overline\psi}^T (x^0,{\bf x},-c) \to \\
\displaystyle
(\gamma^a p_a - mc){\psi}_Q (x^0,{\bf x},-c) = -(e/c)\gamma^a A_a
{\psi}_Q (x^0,{\bf x},-c)
\end{array}
\end{equation}
\noindent Here matrix $U_Q$ satisfies the conditions (\ref{g11}) which
define its  explicit  form  (\ref{g14}).  Taking  into  account  $\psi
_{Q}=\gamma ^2\psi ^* (x^0,{\bf x},-c)$ we can write
\begin{equation}
\label{g26}
(\gamma ^a p_a - mc)\gamma ^2\psi ^* (x^0,{\bf x},-c)=
(-e/c)\gamma ^a A_a \gamma ^2\psi ^* (x^0,{\bf x},-c)
\end{equation}
Similarly to  the charge conjugation,  the equation received coincides
with initial Eq.  (\ref{g21}) for the electric  charge  $-e$  and  the
transformed function $\gamma ^2\psi ^*(x^0,{\bf x},-c)$. In accordance
with formulae (\ref{g19}), (\ref{gg19}) and (\ref{g20}) it is possible
to  admit  that  Eq.  (\ref{g26}) describes a particle with the charge
$-e$, 4-momentum $p=(-p^0,-{\bf p})$ and positive energy   $+{\cal E}$
$(-p^0=(+{\cal E})/(-c),-{\bf p}=(+{\cal E})(-{\bf v})/c^2)$.
\begin{sloppypar}
Thus, the   charge   conjugation  $C$  puts  into  correspondence  the
antiparticle  with  characteristics  $(-e,-m,-p,-{\cal E},c)$     from
hyperplane  $+c$  to  the  particle with characteristics $(e,m,p,{\cal
E},c)$. The $Q$-conjugation puts into correspondence the particle with
characteristics  $(-e,m,-p,{\cal E},-c)$   from hyperplane $-c$ to the
particle with characteristics $(e,m,p,{\cal E},c)$.  It is  seen  that
the particle from hyperplane $-c$ with characteristics $(-e,m,-p,{\cal
E},-c)$ may be the redefined antiparticle with respect to the  initial
particle  from  Eq.  (\ref{g21}).  It  is  the  same particle which we
interpret as the antiparticle on hyperplane $+c$.
\end{sloppypar}
\par As  in  the  case  of  the  $C$-conjugation   \cite{Ber68},   the
$Q$-conjugation  forms  the  symmetry transformation of Dirac equation
(\ref{g21})  for  a   charged   particle   if   the   4-potential   of
electromagnetic field $A$ is transformed to the rule $Q(A)=(-A^0,-{\bf
A})$.

\section{Conclusion}
\label{s6}
\par The  inversion  of the speed of light $x^0 \to x^0,  {\bf x}  \to
{\bf x},c \to  -c$  was  considered  in  the  Maxwell  and  the  Dirac
equations. It  is shown that the charge conjugation $C_e$ in classical
sense  and  the  charge  conjugation  $C$  in  quantum  sense  can  be
interpreted as the consequence of the symmetry of these equations with
respect to the discrete transformation $c \to -c$.
\par Among  consequences  of  the  $c  \to  -c$  symmetry  we can note
the following ones.
\par In  accordance  with classical and quantum electrodynamics we can
admit that the world as  a  whole  is  the  five-dimensional  one.  It
consists   of  two  hyperplanes:  $(x^0,{\bf  x},+c)$  and  $(x^0,{\bf
x},-c)$,  where $c=3.10^{10}$ cm/s.  Each of the hyperplanes forms the
4-dimensional   Minkowski   subspace   with   metric  tensor  $g_{ab}=
diag(+,-,-,-)$ and each of the hyperplanes is filled with photons  and
electrons.
\par The  equations  of  classical and quantum electrodynamics are the
same for $+c$ and $-c$ hyperplanes.  The electron-photon Dirac  vacuum
exists  on  the each hyperplane.  The free photon with positive energy
from $-c$ hyperplane is the same object  as  the  vacuum  photon  with
negative energy from $+c$ hyperplane.  The free electron with positive
energy from hyperplane $-c$ is the same object as the vacuum  electron
with  negative  energy  from  $+c$ hyperplane.  The $-c$ hyperplane is
equivalent to the Dirac vacuum  from  $+c$  hyperplane,  or  the  $-c$
hyperplane show himself as the Dirac vacuum.  Therefore further we can
deal with the Dirac vacuum as with the more known object.
\par With this point of view the reaction $e^-  +  e^+  \to  \gamma  +
\gamma$  can be interpreted as the act of annihilation of the electron
$e^-$ and the  electron  hole  $e^+$  \cite{Sch57}.  Analogously,  the
reaction  $\gamma + \gamma \to e^- + e^+$ we can interpret as the  act
of annihilation of the uncharged photon and uncharged photon hole.  In
the case of the electron vacuum the vacuum transitions $e^- \to {e^-}'
+ \gamma$ are impossible because of the Pauli principle  \cite{Sch57}.
In  the  case  of the photon vacuum the vacuum transitions $\gamma \to
\gamma' + e^- + e^+$ are impossible due to the law of  energy-momentum
conservation,   because   of  the  equations  of  the  energy-momentum
conservation $-\hbar\omega = -\hbar\omega' + {\cal  E}^{(e)}  +  {\cal
E}^{(p)};  \  -\hbar\omega  {\bf n}/c = -\hbar\omega {\bf n}'/c + {\bf
p}^{(e)}  +  {\bf  p}^{(p)}$  are  non-consistent  in  general.   Here
$-\hbar\omega$  and  $-\hbar\omega'$  are the negative energies of the
vacuum  photons  with  the  frequencies   $\omega'>\omega>0$;   ${\cal
E}^{(e)}>0$  and  ${\cal  E}^{(p)}>0$ are the positive energies of the
free electron and positron;  ${\bf n}$ and ${\bf n}'$ are the  guiding
vectors of the vacuum photons; ${\bf p}^{(e)}$ and ${\bf p}^{(p)}$ are
the momenta of the free electron and positron. As this takes place, we
are going from hypothesis that the transitions of vacuum electrons are
accompanying by generation of the free photons and the transitions  of
vacuum   photons   are   accompanying   by   generation  of  the  free
electron-positron pairs or another  pairs  particle-antiparticle  with
total zero charge.
\par  In  the same time spontaneous photon transitions from the states
with positive energy to the states with negative energy  are  possible
as  the reaction of annihilation $\gamma + \gamma \to \nu + {\overline
\nu};  e^- + e^+;  \ldots$.  Products of  annihilation  (neutrino  and
antineutrino,  electron  and positron and etc.) may annihilate in turn
and generate the new gamma quanta.  As a result of this the continuous
process  of energy exchange will be established between the vacuum and
real world,  or in other words,  between the $-c$ hyperplane  and  the
$+c$   hyperplane.   The  electron-photon  vacuum  acquires  dynamical
character.  Perhaps this is just the physical sense of the  "minus  c"
symmetry  in  addition  to  the  interrelation  between  $C$  and  $Q$
conjugations.

\section{Acknowledgments}

The author is deeply grateful to Prof. Y.S. Kim, to  Prof. V.I. Man'ko
and Dr. M.A. Man'ko for attention, discussion of the present work  and 
useful remarks.

\end{document}